\documentclass[12pt,iicol,referee]{article}

\usepackage[font={small}]{caption}
\usepackage{scicite}
\usepackage{physics}
\usepackage{times}
\usepackage{soul}
\usepackage{graphicx}
\usepackage{amsmath}
\usepackage{url}
\usepackage[dvipsnames]{xcolor}

\newcommand{\firstVersion}[1]{}

\newenvironment{sciabstract}{%
\begin{quote} \bf}
{\end{quote}}

\title{Sensing Intelligence as a Trainable Metamaterial Property}

\author
{Kyungmi Na$^{1}$, Yifei Li$^{2}$, Xinyi Yang$^{1}$, Bolei Deng$^{1,\ast}$\\
\normalsize{$^{1}$Daniel Guggenheim School of Aerospace Engineering, Georgia Institute of Technology}\\
\normalsize{$^{2}$ Computer Science and Artificial Intelligence Laboratory, Massachusetts Institute of Technology}\\
\\
\normalsize{$^\ast$Correspondence author. E-mail: bolei.deng@gatech.edu.}\\
}

\date{}

\begin{document} 


\baselineskip24pt


\maketitle 


\begin{sciabstract}
In biological systems, sensing is not performed by the brain alone: the body deforms, vibrates, and filters external stimuli before they are transduced into neural signals. In engineered systems, this processing burden is placed largely on electronics and computation, while the mechanical body is usually designed only for strength and stability. Here, we present sensing intelligence as a trainable property of the body. We show that the geometry of a metamaterial can be optimized to reshape external stimuli into internal signals that are easier for a neural network to interpret. Rather than hand-designing this physical preprocessing, we let the neural network train its own body for sensing by backpropagating the sensing loss to the body's design parameters through differentiable simulation. Across numerical and experimental sensing scenarios, the optimized body improves sensing accuracy by up to fivefold or reduces the number of required electronic sensors by nearly an order of magnitude.
\end{sciabstract}

\section*{Introduction}
\begin{figure}
    \centering
    \includegraphics[width=0.9\linewidth]{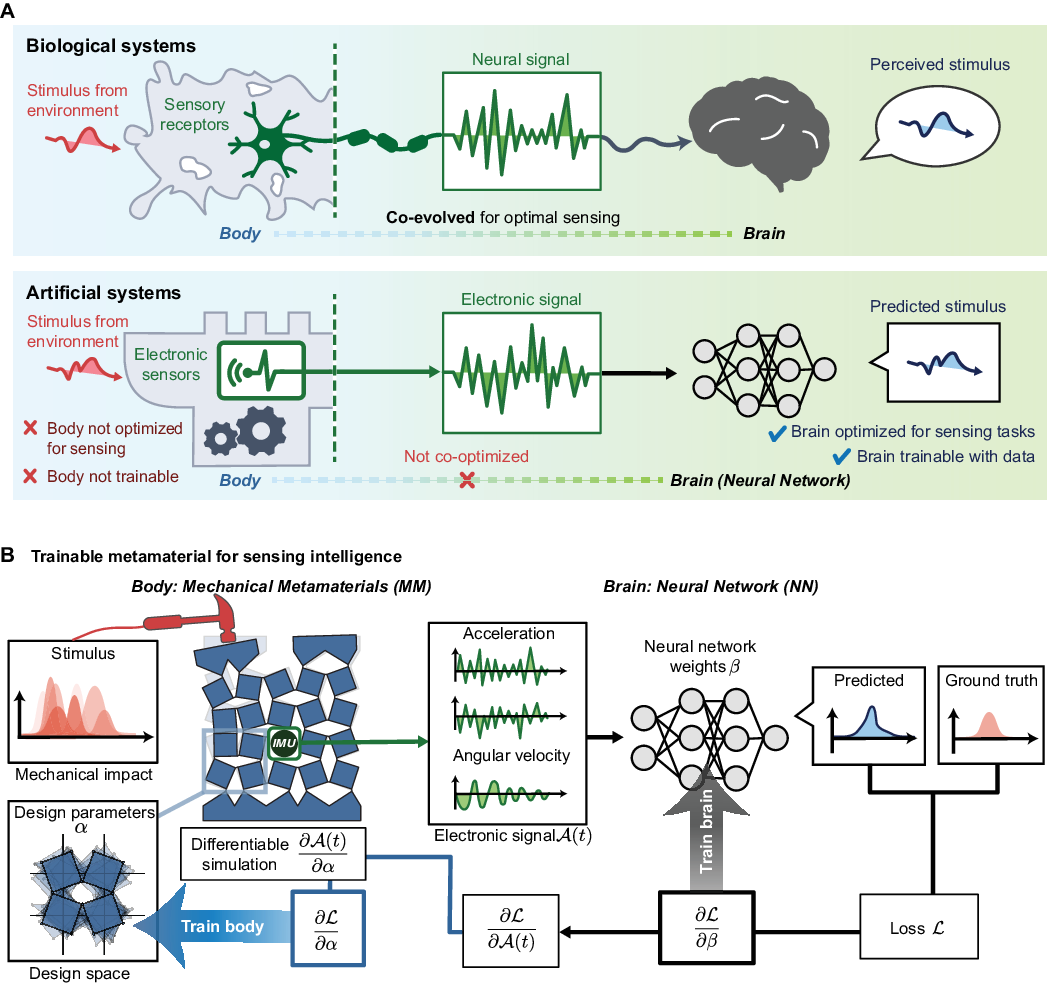}
    \caption{\textbf{General framework for training sensing intelligence in metamaterial bodies.} \textbf{(A)} In biological systems, external stimuli are first transformed by the body into internal physical signals before being interpreted by the brain. In conventional engineered systems, by contrast, the mechanical body is typically fixed and only the computational brain is trained. \textbf{(B)} Our framework treats the body itself as trainable for sensing. A programmable mechanical metamaterial (MM) with design parameters $\alpha$ receives external force stimuli $f_i(t)$ and generates internal IMU signals $\mathcal{A}(t)$, which are decoded by a neural network (NN) with parameters $\beta$. During training, the sensing loss $\mathcal{L}$ backpropagates not only to NN weights via $\partial \mathcal{L}/ \partial {\beta}$, but also through the differentiable metamaterial simulator to the MM body design as $\partial \mathcal{L}/ \partial {\alpha}$, allowing the brain to shape the body into one that produces signals that are easier to interpret.}
    \label{manufigure1}
\end{figure}

The ability to sense and respond to the surrounding environment is fundamental to survival and to nearly all biological activity. Yet sensing does not originate in the brain. Environmental stimuli such as forces, vibrations, and impacts usually first pass through the body, causing tissues to deform or structures to resonate, transforming raw inputs into internal signals that neural systems can interpret. Neural circuits then transmit these body-shaped signals to the brain, where they are further processed to infer what is happening in the external environment. Over evolutionary time, this coupling is refined: morphology, material properties, and neural inference co-adapt so that the signals arriving at the nervous system are not arbitrary, but structured in ways that make downstream interpretation more efficient~\cite{gong_morphological_2023, tiwary_what_2025} (Fig.~\ref{manufigure1}A). Several examples illustrate how morphology facilitates perception, including the whiskers of rats and seals~\cite{towal_morphology_2011, hanke_harbor_2010}, the cochlea of the ear~\cite{mammano_biophysics_1993}, the sensilla morphology of crayfish~\cite{pravin_effects_2015}, and spider webs and legs~\cite{wu_bio-inspired_2026}. Building on these biological insights, engineered robotic systems have increasingly exploited body morphology and mechanical structure as part of the sensing pipeline~\cite{ghazi-zahedi_editorial_2021, pfeifer_soft_2013, nguyen_mechanics_2022, vollrath_spider_2020}. Yet in most artificial systems, the body that shapes incoming stimuli and the neural system that interprets them are still largely designed and optimized in isolation.

In engineered systems, the ``brain'', increasingly realized as a neural network, can be trained end-to-end for perception and decision-making, with parameters updated systematically through gradients and large datasets~\cite{lecun_deep_2015, iskandar_intrinsic_2024, li_developments_2025, li_biomimetic_2026}. By contrast, the ``body'', namely the mechanical structure and material through which environmental stimuli are first received and transformed, is typically regarded as a passive substrate. It is usually designed for strength, packaging, stability, or manufacturability rather than for sensing~\cite{jiaoMechanicalMetamaterials2023}. Even when the body strongly shapes the measured signals, it is rarely treated as a trainable part of perception. If the body can reshape external stimuli into internal signals that are more interpretable to the brain, it becomes an active partner in the sensing process rather than a passive interface. We refer to this capability of the physical body as sensing intelligence. Although recent work has shown that physical systems can be trained to perform computational tasks in ways analogous to neural networks~\cite{wrightDeepPhysicalNeural2022,wetzstein_inference_2020,xue_fully_2024}, our goal here is fundamentally different. We are not trying to substitute any part of computational intelligence, namely the brain. Instead, we train the physical body to accommodate and assist the brain by transforming external stimuli into signals that are easier for it to interpret.

To realize this idea, the physical body must be sufficiently expressive to do more than passively transmit external stimuli. It must possess rich internal dynamics that can filter, rearrange, and redirect those stimuli into signals that are easier for the brain to interpret. Mechanical metamaterials provide a natural platform for this role. Because their effective behavior is governed primarily by geometry, their dynamic response can be systematically programmed through structural design~\cite{bertoldiFlexibleMechanicalMetamaterials2017, kadic3DMetamaterials2019,bordiga_automated_2024}. In this work, we focus on a representative class of metamaterials based on rotating quadrilateral mechanisms, which have been shown to exhibit strong geometric nonlinearity and rich local dynamics~\cite{bordiga_automated_2024, bertoldi_negative_2010, mullin_pattern_2007,deng2017elastic,deng2018metamaterials, coulais_discontinuous_2015,dengCharacterizationStabilityApplication2020, deng2022inverse}. These rich dynamics have been leveraged to realize embodied mechanical computation and control in autonomous robots~\cite{bosch2023autonomous}. Such properties make them well suited for transforming external mechanical inputs into internal signals that support sensing. Furthermore, a differentiable simulator has recently been developed for this class of systems~\cite{bordiga_automated_2024}, enabling precise gradient-based optimization of the dynamic response with respect to intricate geometric parameters. This provides exactly the foundation needed to train the body itself for sensing intelligence.

% Figure 1

\subsection*{General framework for training sensing intelligence}

Here, we present a concrete example to demonstrate the framework for training the body for sensing intelligence (Fig.~\ref{manufigure1}B). Our simple sensing system consists of two coupled components: a physical ``body'', realized as a piece of mechanical metamaterial (MM, with design parameters denoted by $\alpha$), and a computational ``brain'', realized as a neural network (NN, with weights denoted $\beta$). When an external impact force $f_i(t)$ is applied to one of the top panels of the metamaterial body, it excites nonlinear mechanical waves that propagate through the structure. These internal dynamics are captured by an inertial measurement unit (IMU) inside the MM, which records time-series signals including acceleration and angular velocity, denoted by $\mathcal{A}(t)$. The brain then uses these internal signals to infer what happened outside, namely to reconstruct the external impact force $f_i(t)$  from the IMU measurements $\mathcal{A}(t)$ without placing sensors directly on the surface. The details of simulation and design space are given in Supplementary Information S1.

In the training process, the NN is trained in the usual way: its weights $\beta$ are updated through backpropagation to minimize the discrepancy between the predicted and ground truth forces, defined as sensing loss $\mathcal{L}$. Crucially, backpropagation not only provides a gradient to the NN weights via $\partial \mathcal{L}/\partial \beta$ but also produces a gradient of the loss with respect to the input signal, $\partial \mathcal{L}/\partial \mathcal{A}(t)$. In a conventional learning pipeline, this quantity is usually discarded because the input to a NN is usually considered fixed. Here, however, it carries exactly the information we need: it tells us how the brain wants the internal signal to change to predict the external stimulus more accurately. Importantly, in our system the internal signal $\mathcal{A}(t)$ is generated by the MM body, and can therefore be reshaped by changing the body geometry $\alpha$. Because the metamaterial simulator is differentiable, we can compute how the internal signal changes with respect to the body design, namely $\partial \mathcal{A}(t)/\partial \alpha$. This means that the preference expressed by the brain is no longer just a passive gradient at the input. It becomes an actionable training signal for the body itself. By combining these two gradients through the chain rule, we obtain
\[
\frac{\partial \mathcal{L}}{\partial \alpha}
=
\frac{\partial \mathcal{L}}{\partial \mathcal{A}(t)}
\frac{\partial \mathcal{A}(t)}{\partial \alpha}.
\]
This derivation establishes how the metamaterial body, governed by $\alpha$, can be evolved to generate the specific internal signals required for more efficient learning by the brain. In this way, the brain communicates its training preference back to the body, and the body adapts to generate signals that are easier for the brain to interpret. Using this framework, we show that optimized sensing intelligence can improve sensing accuracy by up to fivefold and reduce the number of required electronic sensors by nearly an order of magnitude. More broadly, this establishes a general pathway for co-training bodies and brains, so that physical structure actively supports perception rather than merely serving as a passive interface. 

% Figure 2
\begin{figure}
    \centering
    \includegraphics[width=0.9\linewidth]{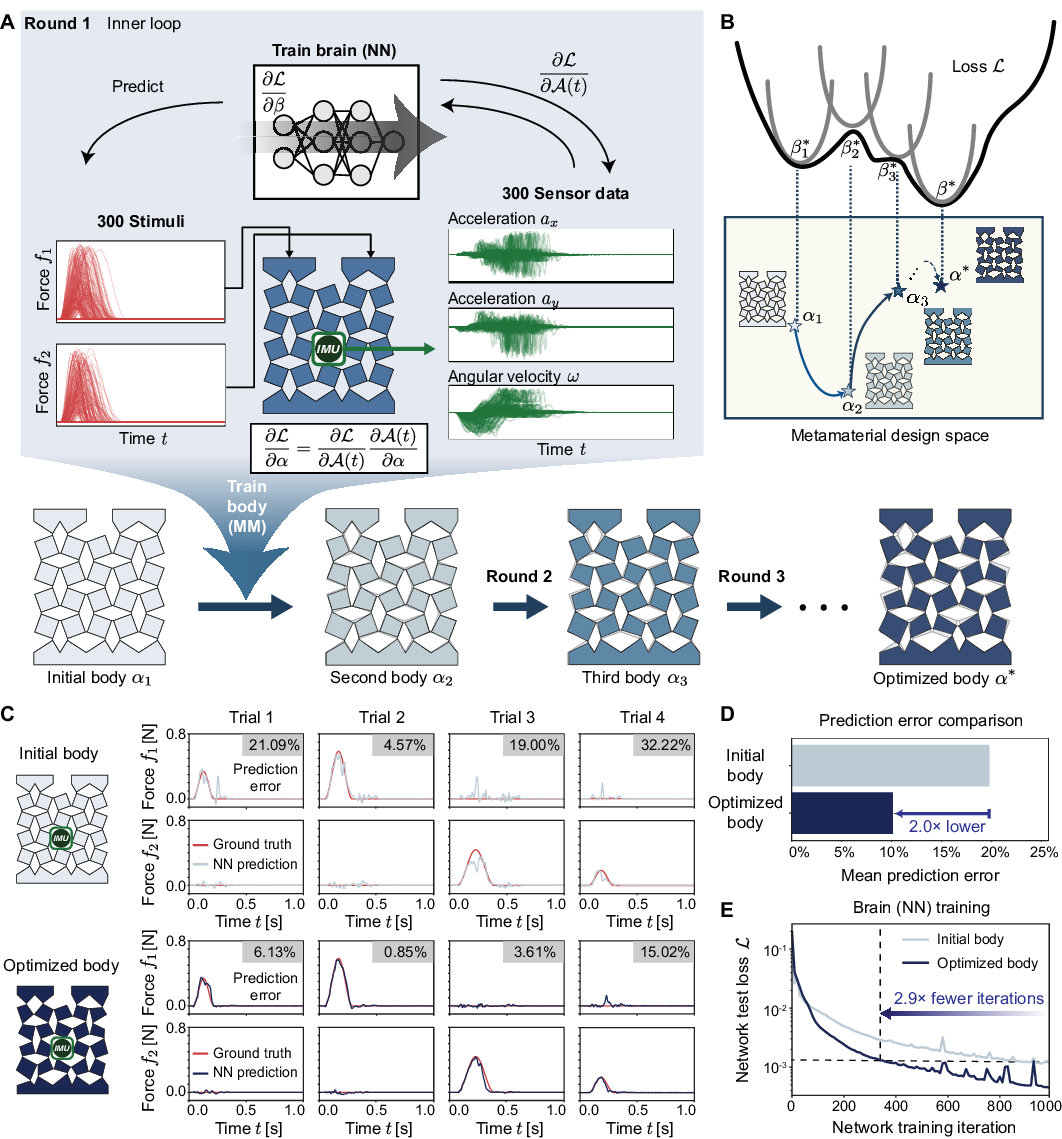}
    \caption{\textbf{Training process and sensing performance of optimized metamaterial bodies.}
\textbf{(A)} Training protocol for the simple sensing system: while the NN is trained to reconstruct the force histories from the IMU measurements, the sensing loss is further backpropagated through the differentiable simulator to update the body geometry.
\textbf{(B)} Nested optimization over body and brain. For each body design $\alpha_i$, the neural network is trained to its corresponding optimum $\beta_i^{*}$, and the resulting sensing gradient is used to update the body toward the next design. Repeating this alternating process drives the system toward an optimized body $\alpha^{*}$ paired with the corresponding optimal brain $\beta^{*}$.
\textbf{(C)} Representative reconstructions of the input force signals for the initial and optimized bodies. The optimized body consistently enables more accurate prediction while using the same NN architecture.
\textbf{(D)} Mean prediction error across all 30 test trials for the initial and optimized bodies.
\textbf{(E)} NN test-loss history during training when paired with the initial and optimized bodies. Signals from the optimized body enable faster learning and reach a lower final loss under the same NN architecture and training budget. }
    \label{manufigure2}
\end{figure}

\subsection*{Training process and sensing performance}

Figure~\ref{manufigure2}A shows the full training protocol for the simplest realization of our framework. We consider a metamaterial body with two input panels on its top boundary that receive external force stimuli, and a single embedded IMU located near the center of the structure. 
To generate the training set, we apply 300 virtual impact signals to the two input panels, denoted by $f_1(t)$ and $f_2(t)$, with different amplitudes, shapes, and time delays. For each set of impacts, the differentiable simulator computes how the triggered nonlinear waves propagate through the metamaterial body and records the acceleration and angular velocity, denoted as $\mathcal{A}(t)=\bigl(a_x(t),a_y(t),\omega(t)\bigr)$ at the IMU location during one second discretized by $T=200$ timesteps. We then train the brain, implemented as a 1D Convolutional Neural Network, to infer the external force history from the internal IMU measurements. The sensing loss $\mathcal{L}$ is defined as the mean squared error between the predicted and ground truth force signals (see Supplementary Information S2 for details)~\cite{kiranyaz_1d_2019}.

% The prediction error is defined as the ratio of the mean squared error to the sum of squared ground truth values, expressed as a percentage, providing a scale-invariant measure of relative reconstruction accuracy}

We first train the brain on the initial body with design parameter set $\alpha_1$, yielding the corresponding optimized NN weights $\beta_1^*$ tailored to this specific MM body. The sensing loss is then backpropagated through the trained NN to obtain its gradient with respect to the IMU signal, $\partial \mathcal{L}/\partial \mathcal{A}(t)$. While this gradient can in principle be obtained via automatic differentiation~\cite{baydinAutomaticDifferentiationMachine2018}, doing so through all network training epochs would cause memory overflow; we therefore employ implicit differentiation to evaluate it efficiently (see Supplementary Information S2.4)~\cite{blondel_efficient_2022,rajeswaran_meta-learning_2019}. Through the differentiable simulator, this signal is further propagated to the body design parameters, yielding the body-training gradient $\partial \mathcal{L}/\partial \alpha$. Again, this gradient carries the brain's preference over the body: it tells us how the MM parameters $\alpha$ should change to produce internal responses that are more informative and easier for the brain to learn from. We then use this gradient to update the MM design, obtaining a second body from the initial one, namely $\alpha_2 = \alpha_1 - \eta \, \partial{\mathcal{L}}/\partial{\alpha}$, where $\eta$ is the learning rate. This process is repeated iteratively. Figure~\ref{manufigure2}B summarizes this nested optimization scheme as a search over the metamaterial design space $\alpha$. For each body design $\alpha_i$, the brain is trained to obtain its corresponding optimal weights $\beta_i^*$ that best accommodate that body. The resulting NN then provides feedback to update its own body design, producing the next body in the sequence. Repeating this alternating process gradually drives the system toward an optimized body design $\alpha^*$ and its corresponding trained brain $\beta^*$. Throughout this process, not only is the brain optimized for its body, but the body is also optimized for its brain, much like in living organisms.

The optimized MM body substantially improves the sensing accuracy of the system. In Fig.~\ref{manufigure2}C, we show four representative cases comparing the force signals predicted by the NN against the ground truth (Movie S1). As expected, the optimized MM body allows the NN to reconstruct the input force signals much more accurately, with consistently smaller relative prediction errors than the initial body. The same advantage holds across all 30 test trials, where the optimized body achieves a markedly lower mean prediction error (Fig.~\ref{manufigure2}D). Notably, both the initial and optimized bodies are paired with the same NN architecture and same training setup. In other words, the brain itself does not become more complex; rather, it is the sensing intelligence of the MM body that improves.

The advantage of the optimized MM body is also evident during the training of the NN brain itself. As shown in Fig.~\ref{manufigure2}E, when the NN is trained on signals generated by the optimized body, its loss decreases much faster than when it is trained on signals from the initial body. To reach the same prediction accuracy, the same NN architecture requires 2.9 times fewer training iterations paired with the optimized body. In other words, a body with improved sensing intelligence reduces the burden on its brain~\cite{sittiPhysicalIntelligenceNew2021}. We further verify that this advantage is not only a transient training-speed effect: after extended training for 10,000 iterations, the NN trained with signals from the optimized body still converges to a 5.8-fold lower prediction error than that from the initial body (see Supplementary Fig.~S2). This captures the essence of sensing intelligence in our framework: the body is trained to reshape external stimuli into a form that is easier for the brain to interpret. As a result, the same neural architecture can achieve better sensing performance with the same computational resources, or equivalently, reach comparable functionality with less computational power. 

% Figure 3
\begin{figure}
    \centering
    \includegraphics[width=0.9\linewidth]{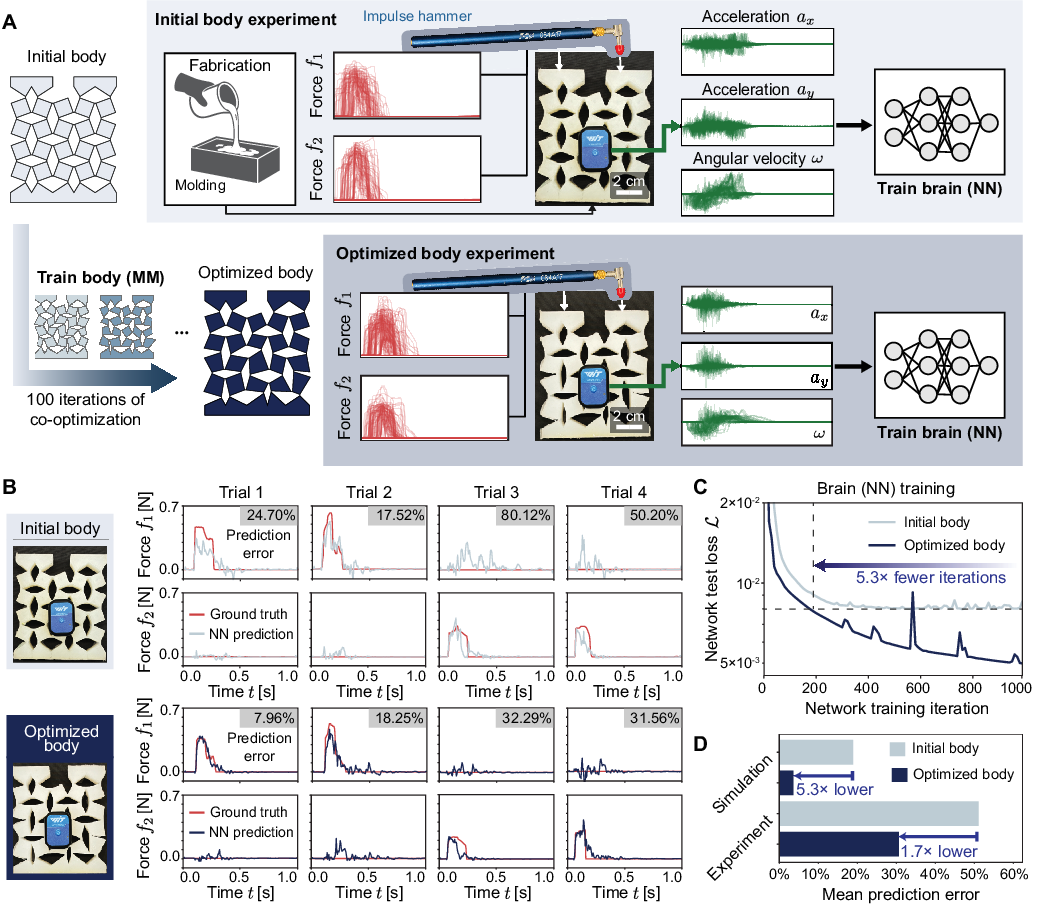}
    \caption{\textbf{Experimental demonstration of sensing intelligence.} \textbf{(A)} The initial MM body is fabricated and tested through impact experiments, where the applied force is measured by an instrumented hammer and the internal response is measured by an embedded IMU. The experimentally measured force signals are then used in the same co-optimization framework with differentiable simulation to update the body geometry. After obtaining the optimized design, the optimized MM body is fabricated and tested again.
    \textbf{(B)} Representative reconstructions of the input force signals for the fabricated initial and optimized bodies. The optimized body enables more accurate prediction while using the same NN architecture.
\textbf{(C)} NN test-loss history from physical experiments with the fabricated initial and optimized bodies. Experimentally measured signals from the optimized body lead to substantially faster learning under the same NN architecture and training budget.
\textbf{(D)} Mean prediction error across all 30 test trials for the initial and optimized bodies for the simulation and experimental cases.}
    \label{manufigure3}
\end{figure}

\subsection*{Experimental demonstration of sensing intelligence}

Having established the framework in simulation, we next evaluate whether sensing intelligence can be realized in the real world. To do so, we fabricate the MM body and replicate the same body--brain pipeline in physical experiments (Fig.~\ref{manufigure3}A). The samples are cast in Dragon Skin\textsuperscript{\textregistered}-10 Fast silicone using 3D-printed molds (see Supplementary Information S3 for details on fabrication, material characterization, and system identification). To measure the external stimulus, we use an instrumented impulse hammer (Model 086E80, PCB Piezotronics), which records the force signal during the moment of impact. Similar to the simulation setup, the internal response is measured using a physical IMU sensor (WT9011DCL, WTMOTION) glued to one of the central quadrilateral blocks of the MM body, recording acceleration and angular velocity during the impact tests. Mechanical impacts are applied manually using an instrumented hammer, which generates and records the applied force history during each impact (see Supplementary Information S3.3 for testing details).

We first fabricate the initial body design and perform 300 random impact experiments on the two input panels. In each trial, an instrumented hammer strikes one panel and records the applied force, while the embedded IMU measures the resulting internal vibration. This yields paired data of external force histories and corresponding internal IMU signals, mirroring the sensing setup in simulation. These data are then used to train the NN to reconstruct the external impact force from the internal body response. Following the same framework introduced above, the trained network provides the gradient of the sensing loss with respect to the IMU signal through backpropagation. From this point onward, we return to the exact same co-optimization pipeline used in simulation: the experimentally measured force signals are used to drive the differentiable simulator, and the body geometry is iteratively updated in simulation until reaching the optimal body design $\alpha^*$. Unlike the idealized bell-shaped pulses used in Fig.~\ref{manufigure2}, the experimentally measured force profiles are much more irregular and less smooth, making the sensing task substantially more challenging. In this more unstructured real-world sensing setting, body optimization becomes even more critical: the optimized body reduces the relative prediction error by 5.3-fold compared with the initial body (see Supplementary Fig.~S5), a much larger gain than the twofold improvement observed for the idealized bell-shaped pulse setting in Fig.~\ref{manufigure2}.
%This shows that training the body for the brain is particularly valuable in unstructured real-world sensing scenarios.

Upon obtaining the optimized design, we fabricate the optimized MM body using the same procedure and repeat the impact experiments to collect a new dataset for NN training. The optimized body again enables the same NN architecture to predict external force stimuli more accurately and to train more easily than the initial body. Since it is impossible to generate identical impulse inputs across separate experiments, we select four representative trials with similar measured input force profiles for the initial and optimized bodies. As shown in Fig.~\ref{manufigure3}B, under these comparable experimental inputs, the fabricated optimized body clearly outperforms the initial body in reconstructing the external force stimuli (Movie S2). Note that because the NN architecture is identical in both systems, this improvement does not come from a stronger brain. Instead, it comes from the increased sensing intelligence of the body, which generates internal signals that are easier for the same brain to learn. As a result, the training loss of the NN drops substantially faster when the internal signals are generated by the optimized body (Fig.~\ref{manufigure3}C). To reach the same prediction accuracy, the same NN architecture requires 5.3$\times$ fewer training iterations with the optimized body.

Across all 30 real-world impulse tests, the optimized body reduces the mean prediction error by 1.7-fold compared with the initial body (Fig.~\ref{manufigure3}D). The experimentally observed improvement in sensing performance is smaller than that predicted by simulation, as expected given several real-world uncertainties: our idealized discrete model is less accurate for high-amplitude dynamic impulses (Supplementary Fig.~S4D), the IMU measurements are noisier, and the impulse inputs applied to the initial and optimized bodies cannot be made identical. Nevertheless, despite these simulation-to-reality discrepancies, the optimized body still consistently improves sensing performance. This indicates that the sensing gradient from the brain (NN) provides a useful geometric direction for improving the body, even when the simulation does not capture every physical detail. We further evaluate randomly generated MM geometries in simulation and find that they perform worse than the initial design (Supplementary Information S4), confirming that the improvement arises from the optimized geometry rather than arbitrary geometric variation. Together, these results demonstrate that sensing intelligence is a robust physical property of the optimized body rather than an artifact of simulation.

\begin{figure}
    \centering
    \includegraphics[width=0.9\linewidth]{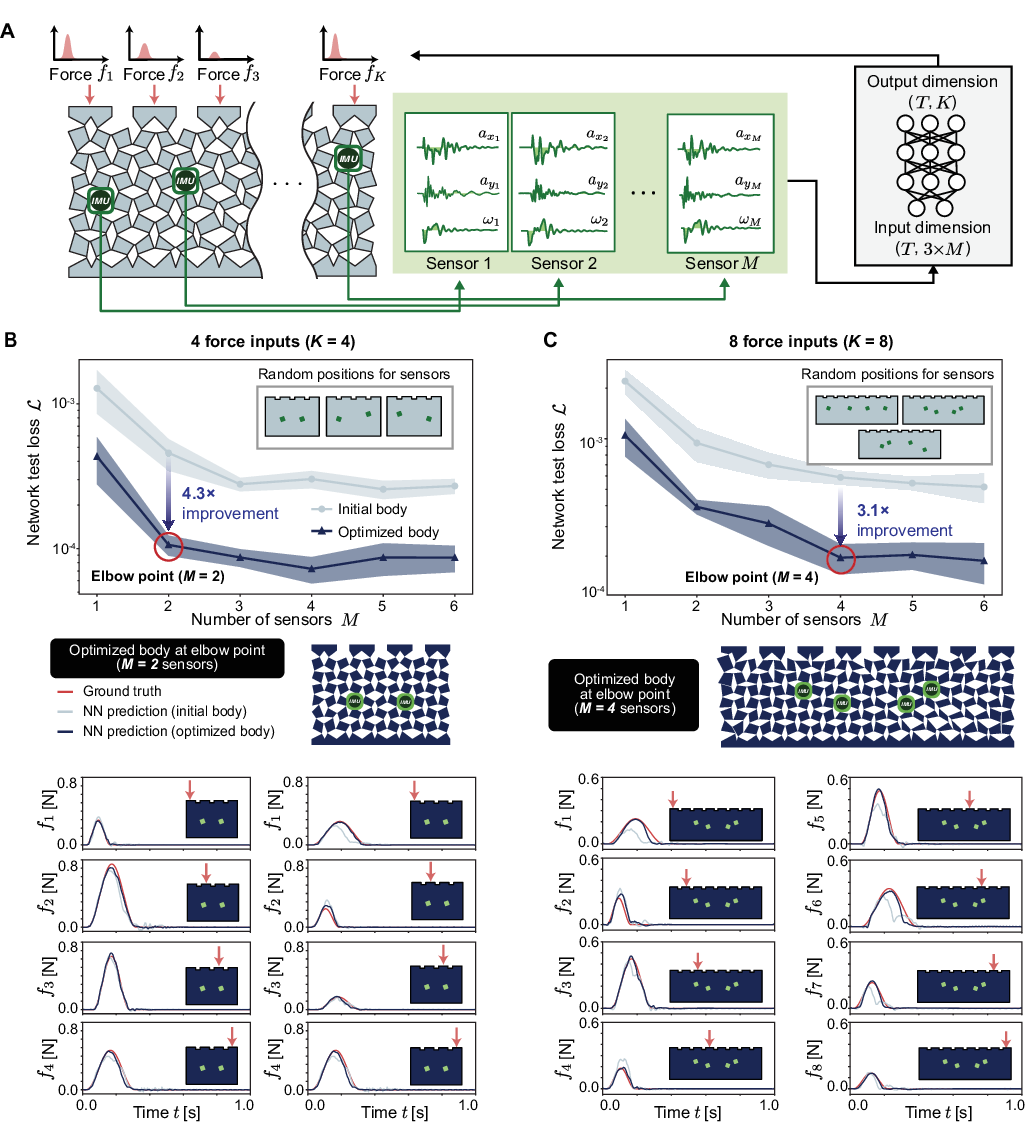}
   \caption{\textbf{Generalized sensing with multiple inputs and sensors.} \textbf{(A)} General sensing setting with $K$ input panels for applying external force stimuli and $M$ embedded IMU sensors for measuring internal mechanical signals. Signals from all $M$ sensors are concatenated and fed into a single NN to reconstruct the force histories applied to all $K$ input panels. \textbf{(B)} Four-input sensing task with $K=4$. The optimized body consistently outperforms the initial body across all tested numbers of sensors. Shaded regions indicate variation across sensor-location configurations. An elbow point appears at $M=2$, beyond which performance saturates. \textbf{(C)} More challenging eight-input sensing task with $K=8$. The elbow point shifts to $M=4$, indicating that the harder task requires more internal sensing capacity before performance saturates.}
    \label{manufigure4}
\end{figure}

\subsection*{Generalized sensing with multiple inputs and sensors}

We next investigate how sensing intelligence scales with task complexity by extending the framework to more general multi-input, multi-sensor systems. Specifically, we consider an MM body with $K$ input panels for external force stimuli and $M$ embedded IMU sensors for internal signal collection (Fig.~\ref{manufigure4}A). The signals from all $M$ sensors are concatenated and fed into a single NN that reconstructs the force histories applied to all $K$ input panels. This setup creates a fundamental trade-off: while additional sensors provide richer information, they also increase the demands on electronics and downstream processing. We hypothesize that a body with higher sensing intelligence can mitigate this computational burden by reducing the number of sensors required to achieve similar sensing performance.

Using the same optimization pipeline, we study two representative cases: MM bodies with four inputs ($K=4$) and eight inputs ($K=8$). For each case, we vary the number of sensors $M$, train each MM body, and compare the initial body with the optimized body. Furthermore, to account for the effect of sensor placement, we test five different sensor-location configurations for each value of $M$ (see Supplementary Information S5). As shown in Fig.~\ref{manufigure4}B, for the $K=4$ case, the optimized body consistently outperforms the initial body across all sensor counts. Interestingly, the optimized body exhibits a clear elbow point at $M=2$, beyond which sensing performance saturates. In fact, the optimized body with only two sensors surpasses the performance of the initial body equipped with six sensors. This suggests that, once the body's sensing intelligence is sufficiently enhanced, substantially less electronic sensing infrastructure is needed to achieve the same or even better functionality. Another interesting observation is that, in this scenario, specific sensor placements do not matter much either. This is because, through the sensing intelligence training process, the body itself evolves around the sensor locations to direct more informative signals toward them.

A similar trend appears in the more challenging $K=8$ input scenario shown in Fig.~\ref{manufigure4}C, where the MM body is much larger and the brain must reconstruct a greater number of external force inputs at every time point. As expected, one or two sensors are no longer sufficient. The elbow point for this scenario shifts to $M=4$, indicating that this harder sensing task requires greater internal sensing capacity before performance saturates. Nevertheless, the optimized body still substantially outperforms the initial body across all sensor counts. The representative trials shown at the bottom of Fig.~\ref{manufigure4}B and C further illustrate that, for both $K=4$ and $K=8$, the optimized body enables much more accurate reconstruction of the external force signals (Movies S3 and S4). It is important to note that, in each comparison, the initial and optimized bodies are paired with the same neural network architecture, so the brain itself does not become more complex. Instead, the body becomes ``smarter,'' enabling the brain to achieve the same function with less hardware, or better performance with the same hardware.

\begin{figure}
    \centering
    \includegraphics[width=0.75\linewidth]{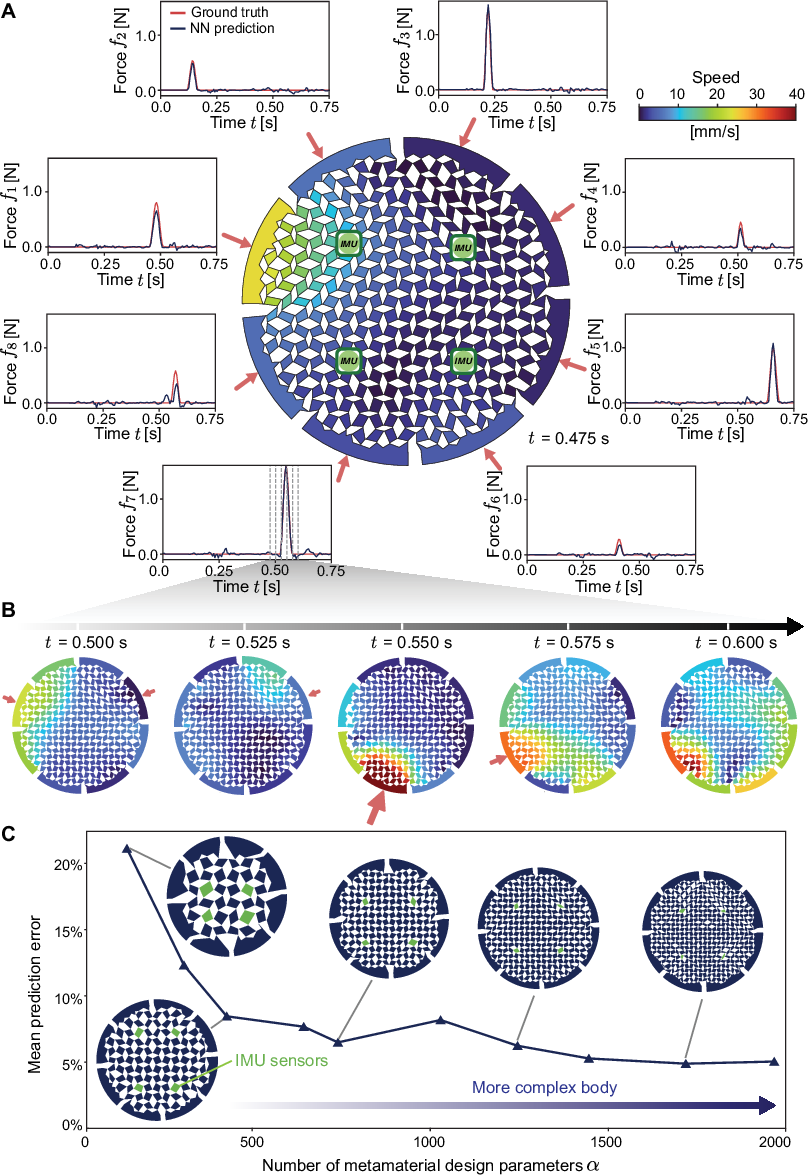}
    \caption{\textbf{Omnidirectional sensing with a circular metamaterial body.} \textbf{(A)} Circular MM body with eight boundary input panels and four embedded IMU sensors. External forces can arrive from different directions, with multiple panels excited within the same sensing window. The optimized body accurately reconstructs force signals from all directions. \textbf{(B)} Speed snapshots showing nonlinear waves propagating through the MM body after impacts from different directions. \textbf{(C)} Effect of body complexity on sensing performance. Increasing the number of trainable design parameters generally improves prediction accuracy while keeping the sensing task, sensor count $(M=4)$, and NN architecture fixed.}
    \label{manufigure5}
\end{figure}

\subsection*{Omnidirectional sensing with a circular body}

In biological organisms and robotic systems, body geometry and sensing scenarios are shaped by function and environment rather than confined to a single layout. To demonstrate that sensing intelligence is not restricted to the particular body configuration studied above, we next consider a compact circular MM body with eight input panels distributed around its boundary, allowing stimuli to arrive from all directions. This provides a more general sensing setting and is relevant to systems such as mobile robotic agents or omnidirectional sensing probes. Based on our previous study of the $K=8$ sensing task, we use $M=4$ embedded IMU sensors to provide sufficient sensing and computational capability (see Supplementary Information S6). We then apply the same co-training framework to this circular system, jointly training the NN brain and the MM body for sensing intelligence.

This omnidirectional sensing task is significantly more demanding than the previous planar configuration. External forces can arrive from all directions, multiple panels are excited at the same time, and each impact can have its own amplitude, sharpness, and temporal profile. Despite this increased complexity, the optimized circular MM body, together with its NN brain, accurately reconstructs all eight external events, as shown in Fig.~\ref{manufigure5}A. The underlying mechanism is illustrated in Fig.~\ref{manufigure5}B: impacts arriving from different directions generate nonlinear waves that propagate throughout the entire body, as reflected by the speed (i.e. magnitude of velocity of quadrilaterals) of individual quadrilateral elements (Movie S5). Through co-training, the MM body learns to shape this wave propagation so that, by the time the resulting internal signals reach the IMU sensors, they carry the most useful information for the brain to infer the external events.

This circular body setting also allows us to examine a central aspect of sensing intelligence: how the geometric complexity of the body affects sensing performance. For the brain, greater capability usually comes from greater computational complexity, for example more sensors or larger neural networks. Here, the body has an analogous form of physical complexity, measured by the number of trainable design parameters in the MM, $\alpha$. To isolate its effect, we keep the sensing task, the sensor count ($M=4$), and the NN architecture fixed, while varying only the geometric complexity of the body by increasing the number of quadrilateral units, thereby increasing the dimensionality of $\alpha$. As shown in Fig.~\ref{manufigure5}C, for a fixed brain architecture and sensing task, prediction performance improves as the body becomes more geometrically complex. This suggests that the body possesses its own form of model capacity: by increasing its geometric expressivity, the body can realize richer mechanical transformations and generate more informative internal signals for sensing. Just as a larger neural network can accomplish more complex tasks, a more complex MM body can perform more sophisticated mechanical transformations and thereby support better sensing. This suggests that intelligence is not only a property of the brain, but also a property that can be scaled within the physical body itself through geometric expressivity.

\section*{Outlook}

In summary, we present sensing intelligence as a trainable metamaterial property. Through body-brain co-optimization, external stimuli are mechanically transformed into internal signals that are easier for a neural network to interpret. Across both simulation and experiment, this co-training framework substantially improves sensing performance, increasing accuracy by up to fivefold or reducing the required number of electronic sensors by nearly an order of magnitude. More broadly, we show that sensing intelligence is not restricted to a particular body geometry or sensing scenario, but extends to more complex layouts and omnidirectional tasks. We further find that, analogous to computational complexity in the brain, geometric complexity in the body can also improve sensing performance, suggesting that intelligence can be scaled through the complexity of the physical body itself.

Another interesting perspective is that our system closely resembles an autoencoder. The MM body acts as a physical encoder that compresses high-dimensional external stimuli into lower-dimensional internal signals, while the neural network acts as a decoder that reconstructs the original stimuli using only those internal signals. The measured acceleration and angular velocity therefore play the role of a latent representation: compact, yet still sufficiently informative for decoding. As in learned autoencoders, the encoder and decoder must be optimized jointly so that the latent representation is not only low-dimensional, but also informative and recoverable. The idea of a physical encoder is also closely related to physical reservoir computing, where the rich dynamics of a physical substrate are harnessed for information processing. In conventional reservoir computing, however, the physical reservoir is often treated as a fixed complex dynamical system, and only the readout layer is trained~\cite{TANAKA2019100, cucchi_hands-reservoir_2022, yan_emerging_2024}. Yet complexity alone does not guarantee useful computation; the reservoir must generate internal states that are complex in the right way for the target task. Our framework provides a way to train the reservoir itself by backpropagating the task loss from the readout NN into the physical body. In this sense, a physical reservoir can be viewed as a trainable physical representation model: it can be optimized to produce internal dynamics that are not only rich, but also informative, robust, and easy to decode. Once trained, such a reservoir could support different downstream sensing tasks by training only lightweight readout NNs on top of its physical latent representations.

More broadly, although our demonstrations focus on a specific type of MM, the framework is not restricted to this architecture. The same principle can extend to other engineered systems whenever the physical response is sufficiently high-dimensional and nonlinear, and a differentiable physical model is available to backpropagate sensing loss to the material design parameters, such as fluids~\cite{li2022anisotropicStokes}, deformable objects~\cite{du2021_diffpd}, articulated bodies~\cite{Xu-RSS-21}, and microstructures~\cite{diffmicrostructure}. In such systems, a parameterized physical body can be trained jointly with a neural network brain so that the body itself carries out part of the information processing required for sensing. For instance, magnetically responsive metamaterials may be designed to encode external magnetic fields into internal mechanical responses that can then be decoded by a learned model. Beyond sensing intelligence, our framework is readily applicable to the broader concept of physical intelligence, in which the body is trained to serve as a synergistic partner to the brain, not only for sensing but also for actuation, coordination, and even computation~\cite{sittiPhysicalIntelligenceNew2021}.

% More generally, our findings point to a route toward morphology--inference co-optimization across robotics, materials science, and structural engineering, where physical form is designed not only for load bearing, motion, or adaptation, but also for perception itself.

\section*{Methods}

\subsection*{The body: mechanical metamaterials}

The physical bodies in our framework are realized as rotating quadrilateral mechanical metamaterials (MMs). The geometry of each rotating quadrilateral system is fully characterized by the positions of its vertex nodes, which together define the continuous design space of the MM, i.e., $\alpha$. This platform is well suited for learning-based sensing for three reasons. First, it generates rich internal dynamic responses with strong nonlinearity and fading memory, thereby performing a complex spatiotemporal transformation from incoming force histories to internal motion signals. Second, its design space is compact and continuous, enabling efficient optimization over geometric parameters. Third, these design parameters strongly and differentiably influence the resulting dynamics, allowing sensing performance to be improved through gradient-based body optimization.

In a rotating quadrilateral system, the quadrilateral units are modeled as rigid bodies, while the thin ligaments connecting adjacent units are modeled as linear elastic springs that capture stretching, shearing, and bending deformation. The differentiable simulation of this system therefore solves the discrete-time dynamics of the quadrilateral structure by numerically integrating the equations of motion derived from the elastic energy stored in the connecting ligaments. For a system of $N_q$ rigid quadrilateral units and $N_l$ elastic ligaments, the kinetic energy of the $j$-th rigid unit is defined as
\begin{equation}
    T_j = \frac{1}{2}\left[
    m_j\left(\dot{u}_j^2+\dot{v}_j^2\right)
    + I_j\dot{\theta}_j^2
    \right],
\end{equation}
where $u_j$, $v_j$, and $\theta_j$ represent the translational displacements and rotation of the $j$-th unit, while $m_j$ and $I_j$ denote its mass and rotational inertia, respectively. The elastic strain energy of the $i$-th ligament is defined as
\begin{equation}
    U_i = \frac{1}{2}\left[
    k_l \Delta l_i^2
    + k_s \Delta s_i^2
    + k_{\theta}\Delta\theta_i^2
    \right] + E_i^c,
\end{equation}
where $k_l$, $k_s$, and $k_{\theta}$ are the stiffnesses associated with longitudinal, shear, and rotational deformation modes, respectively. The variables $\Delta l_i$, $\Delta s_i$, and $\Delta \theta_i$ denote the corresponding longitudinal, shear, and rotational deflections of the ligament. The term $E_i^c$ is an optional differentiable contact energy used when contact handling is enabled. Summing over all quadrilateral units and elastic ligaments, the total kinetic and elastic energies are given by
\begin{equation}
    T = \sum_{j=1}^{N_q} T_j,
    \qquad
    U = \sum_{i=1}^{N_l} U_i.
\end{equation}
The Lagrangian of the system is then defined as
\begin{equation}
    L = T - U.
\end{equation}
Applying the Euler--Lagrange equation yields the governing equations of motion:
\begin{equation}\label{eq:dynamics}
    \dv{t}\pdv{L}{\dot{\mathbf{s}}}
    - \pdv{L}{\mathbf{s}}
    =
    \mathbf{f}_{\mathrm{ext}}
    +
    \mathbf{f}_{\mathrm{damp}},
\end{equation}
where $\mathbf{s}=[u_1,v_1,\theta_1,\ldots,u_{N_q},v_{N_q},\theta_{N_q}]^{T}$ is the generalized state vector containing the translations and rotations of all rigid units. The vectors $\mathbf{f}_{\mathrm{ext}}$ and $\mathbf{f}_{\mathrm{damp}}$ represent the generalized external forces and viscous damping forces, respectively. Although the simulator supports polygonal structures with an arbitrary number of nodes, all optimized designs in this study are restricted to quadrilateral units with fixed connectivity. This ensures that the optimization explores continuous geometric variables, namely node positions, without requiring topological changes. The design space, strain-energy formulation, governing equations, and numerical integration scheme are detailed in Supplementary Information S1.

The differentiable simulator is implemented in JAX for high-performance automatic differentiation \cite{bradbury_jax_2018}. Numerical integration is performed using a fifth-order explicit Runge--Kutta method from the Diffrax library \cite{kidger2021on, tsitouras2011runge}. Unless otherwise specified, each simulation is conducted over a duration of 1~s and sampled at 200 equally spaced temporal measurement points. This time window can be adjusted to match the characteristic response time of different physical systems.

\subsection*{The brain: neural network}

Within the proposed framework, neural networks (NNs) reconstruct external force time series from internal motion measurements — specifically, translational acceleration and angular velocity. This constitutes a temporal regression problem in which the inputs are sequences of internal sensor measurements and the outputs are sequences of external force signals. We employ a one-dimensional convolutional neural network (1D CNN), which is well suited to extracting temporal features from sequential data~\cite{kiranyaz_1d_2019}. Unless otherwise specified, the baseline model predicts two force channels from a single IMU that provides three input features: the $x$-axis and $y$-axis accelerations and the angular velocity. For systems with multiple sensors or more force inputs, the input and output dimensions are adjusted accordingly while the network architecture is kept the same.

\paragraph{Network architecture}
The baseline 1D CNN architecture is summarized in Table~\ref{tab:timeseriescnn}. It consists of two convolutional layers with 64 filters and kernel size $k=3$, each followed by a Rectified Linear Unit (ReLU) activation. A final convolutional layer with kernel size $k=1$ maps the latent temporal features to the output force channels. All convolutional layers use stride 1 and \texttt{SAME} padding so that the temporal length is preserved throughout the network. The network therefore produces a force prediction at every time step.

For a system with $M$ sensors, the input has $3M$ channels because each sensor measures two translational accelerations and one angular velocity. For a system with $K$ external force inputs, the final convolutional layer has $K$ output channels. The number of trainable parameters is therefore
\begin{equation}
    \text{Number of parameters} \;\beta = (3 \cdot 3M \cdot 64 + 64)
    + (3 \cdot 64 \cdot 64 + 64)
    + (1 \cdot 64 \cdot K + K),
\end{equation}
where the three terms correspond to the first, second, and final convolutional layers, respectively. We intentionally use a lightweight network with approximately $10^4$ trainable parameters so that improvements in sensing performance primarily reflect the optimized physical body rather than the capacity of the neural decoder.

\begin{table}[tbh!]
\centering
\caption{Architecture of the 1D CNN for time-series force reconstruction. All convolutions use stride 1 and \texttt{SAME} padding, so the temporal length $T$ is preserved. $M$ denotes the number of sensors and $K$ denotes the number of external force inputs. The baseline setting uses $M=1$ and $K=2$.}
\label{tab:timeseriescnn}
\begin{tabular}{l l c c}
\hline
\textbf{Layer} & \textbf{Operation} & \textbf{Kernel} & \textbf{Output shape} \\
\hline
Input & --- & --- & $(T, 3M)$ \\
Conv1 + ReLU & 1D convolution, $C_{\mathrm{out}}=64$ & $k=3$ & $(T, 64)$ \\
Conv2 + ReLU & 1D convolution, $C_{\mathrm{out}}=64$ & $k=3$ & $(T, 64)$ \\
Conv3 & 1D convolution, $C_{\mathrm{out}}=K$ & $k=1$ & $(T, K)$ \\
\hline
\end{tabular}
\end{table}

\subsection*{Nested optimization of brain and body}

The proposed framework employs a nested optimization procedure consisting of an inner loop that trains the NN parameters and an outer loop that updates the MM design parameters. The inner loop optimizes the NN weights $\beta$ for a given body design $\alpha$, while the outer loop updates $\alpha$ using the sensing gradient propagated from the trained NN back through the mechanical body. Both loops use the Adam optimizer for its efficiency and robustness on nonconvex problems~\cite{kingma_adam_2015}. The learning rates are set to $10^{-2}$ for the inner loop and $10^{-1}$ for the outer loop. Unless otherwise specified, the NN is trained for 1,000 iterations per outer-loop step, and the MM design parameters are updated over 100 outer-loop iterations.

The loss function for body training is the mean squared error (MSE) of the NN predictions evaluated on the held-out test dataset (see Eq.~\eqref{eq:loss}). During NN training, the loss is computed using normalized force values from the training set only.
\begin{equation}
\label{eq:loss}
    \mathcal{L}
    =
    \frac{1}{NKT}
    \sum_{i=1}^{N}
    \sum_{j=1}^{K}
    \sum_{t=1}^{T}
    \left(
    f_{j,\mathrm{true}}^{i}(t)
    -
    f_{j,\mathrm{NN}}^{i}(t)
    \right)^2
    =
    \mathrm{MSELoss}
    \left(
    f_{\mathrm{true}},
    f_{\mathrm{NN}}
    \right),
\end{equation}
where $N$ denotes the number of samples, $K$ denotes the number of target force channels, and $T$ denotes the temporal sequence length. For the baseline two-panel MM case, we use $N=300$, $K=2$, and $T=200$. In studies involving four and eight force inputs, $K$ is set to 4 and 8, respectively. When the loss is evaluated on the training dataset, it is denoted as $\mathcal{L}_{\mathrm{train}}$ and used for inner-loop NN training. When evaluated on the test dataset, it is denoted as $\mathcal{L}_{\mathrm{test}}$ and used as the outer-loop objective for optimizing the physical body.
Additionally, a prediction error is defined as follows:
\begin{equation}
    \text{Prediction error}_i = \frac{ \sum_{j}^{K} \sum_{t}^{T}(f_{j,true}^{i}(t) - f_{j,NN}^{i}(t))^2}{\sum_{j}^{K} \sum_{t}^{T}(f_{j,true}^{i}(t))^2} \times 100 \; (\%)
\end{equation}
The mean prediction error is obtained by averaging the per-sample prediction errors over the test set.

\subsection*{Brain-to-body gradient propagation}

The proposed framework requires propagating the sensing loss back to the material design parameters, denoted by $\alpha$. In standard NN training, gradients with respect to network weights are efficiently computed using automatic differentiation, commonly referred to as backpropagation \cite{baydinAutomaticDifferentiationMachine2018}. Here, however, the goal is not only to update the NN weights, but also to update the geometry of the mechanical body. This requires computing how the optimized NN weights depend on the internal motion signals generated by the body, and how those signals depend on the body design.

Let $\beta^\star$ denote the optimized NN parameters obtained after training, and let $A(t;\alpha)$ denote the time-series IMU measurements generated by differentiable simulation for a body design $\alpha$. The sensing loss on the test set is denoted by $\mathcal{L}_{\mathrm{test}}$ (see Eq.~\eqref{eq:loss}). The gradient of the test loss with respect to the body design can be written through the chain rule as
\begin{equation}
    \pdv{\mathcal{L}_{\mathrm{test}}}{\alpha}
    =
    \pdv{\mathcal{L}_{\mathrm{test}}}{\beta^\star}
    \pdv{\beta^\star}{A}
    \pdv{A(t;\alpha)}{\alpha}.
\end{equation}
The first term, $\partial \mathcal{L}_{\mathrm{test}}/\partial \beta^\star$, is obtained by standard NN backpropagation. The last term, $\partial A(t;\alpha)/\partial \alpha$, is provided by the differentiable simulator, which relates changes in body geometry to changes in the simulated sensor measurements. The middle term, $\partial \beta^\star/\partial A$, is the most challenging component because $\beta^\star$ is itself the result of an iterative NN training process. Directly differentiating through the full training trajectory would require storing and backpropagating through all optimization steps, which is computationally expensive and can lead to memory overflows.

To avoid differentiating through the entire training history, we use implicit differentiation. This approach is well suited for computing gradients through iterative processes that converge to an optimum or fixed point \cite{rajeswaran_meta-learning_2019, blondel_efficient_2022}. We formulate NN training as an optimization problem,
\begin{equation}
    \beta^\star(A)
    =
    \arg\min_{\beta}
    \mathcal{L}_{\mathrm{train}}(\beta, A),
\end{equation}
where $\mathcal{L}_{\mathrm{train}}$ is the training loss of the NN. At convergence, the optimized weights satisfy the first-order optimality condition
\begin{equation}
    F(\beta^\star(A),A)
    =
    \nabla_{\beta}\mathcal{L}_{\mathrm{train}}(\beta^\star(A),A)
    =
    0.
\end{equation}
Applying the implicit function theorem to this optimality condition gives
\begin{align}
    & \pdv{F(\beta^\star(A),A)}{\beta}
    \pdv{\beta^\star(A)}{A}
    +
    \pdv{F(\beta^\star(A),A)}{A}= 0, \\
    & \pdv{F(\beta^\star(A),A)}{\beta}
    \pdv{\beta^\star(A)}{A}
    =
    -
    \pdv{F(\beta^\star(A),A)}{A}.
\end{align}
Therefore, $\partial \beta^\star(A)/\partial A$ can be obtained by solving the corresponding linear system,
\begin{equation}
    \pdv{\beta^\star(A)}{A}
    =
    -
    \left[
    \pdv{F(\beta^\star(A),A)}{\beta}
    \right]^{-1}
    \pdv{F(\beta^\star(A),A)}{A}.
\end{equation}
This implicit formulation allows us to compute the dependence of the trained NN on the body-generated sensor signals without storing the full NN training trajectory. For implicit differentiation, the resulting linear systems are solved using a conjugate gradient solver with a tolerance of $10^{-7}$ and a maximum of 20 iterations.

Combining the NN gradient, the implicit gradient through the trained weights, and the differentiable simulation gradient yields $\partial \mathcal{L}_{\mathrm{test}}/\partial \alpha$. This gradient provides a direct signal for how the body geometry should change to improve sensing performance. In this sense, the trained NN acts as a ``brain'' that sends task-level gradient information back to the mechanical ``body'', allowing the MM geometry to evolve toward more informative internal dynamics.

\subsection*{Fabrication and experiment}
Physical samples were fabricated by casting. Given a set of design parameters, MATLAB code generated an AutoCAD script file. Based on the AutoCAD script, we generated parts of a mold for casting. A Bambu printer (LAB P1S) printed the mold by exporting STL files of the mold to G-code. We used PLA filaments for fabrication of the mold. The two materials of Dragon Skin\textsuperscript{\tiny\textregistered}-10 Fast silicone were poured into the container and mixed by a centrifugal mixer (THINKY ARE-310). The mixed materials were poured into the mold, which had been coated with demolding spray. The samples were cured for 24 hours at room temperature.

To measure forces and physical responses induced by forces, we used an instrumented impulse hammer (Model 086E80, PCB Piezotronics) and an IMU (WT9011DCL, WTMOTION) to record acceleration and angular velocity. During data collection, the experimenter applied impacts to the fabricated samples using the impulse hammer. Prior to co-optimizing the NN and MM, we performed static and dynamic tests to identify simulation parameters. The parameters were chosen to align simulation data with experimental data. After system identification, we applied 150 impacts to each panel, producing 300 input–output pairs for the two-panel configuration. The recorded signals were then preprocessed and used for co-optimization. Details of fabrication, system identification and preprocessing are provided in the Supplementary Information S3.

\section*{Supplementary Information}
We will release the Supplementary Information along with the publication.

\section*{Data availability}
The datasets are available from the corresponding author upon reasonable request.

\section*{Code availability}
The code is available from the corresponding author upon reasonable request.

\section*{Acknowledgments}
B.D. acknowledges support from the NSF Award [IIS-2435905]. The authors thank Giovanni Bordiga for helpful technical assistance in setting up the differentiable simulator used in this work.

% \section*{Author contributions}

\section*{Competing interests}
The authors declare no conflict of interest.

\bibliographystyle{Science}
\bibliography{scibib}

\clearpage

\end{document}